\documentclass[12pt]{iopart}
\usepackage{amsfonts}
\usepackage[english]{babel}
       \newcommand{\Cc}{\mathcal{C}}
       \newcommand{\Dc}{\mathcal{D}}
       \newcommand{\Ec}{\mathcal{E}}
       \newcommand{\Gc}{\mathcal{G}}
       \newcommand{\Hc}{\mathcal{H}}
       \newcommand{\Ic}{\mathcal{I}}
       \newcommand{\Rc}{\mathcal{R}}
       
       \newcommand{\R}{\mathbb{R}}
       \newcommand{\D}{\mathbb{D}}
       \font\bfgreek=cmmib10 scaled 1200
       \newcommand{\bftheta}{\mbox{\bfgreek\symbol{'022}}}
       \newcommand{\w}{\wedge}
       \newcommand{\wt}{\widetilde}
\begin{document}

\title{Contact Equivalence Problem for Linear Hyperbolic Equations}

\author{Oleg I. Morozov}

\address{Department of Mathematics, Snezhinsk Physical and Technical Academy,

\noindent
Snezhinsk, 456776, Russia

\noindent
oim{\symbol{64}}foxcub.org}

\begin{abstract}
We consider the local equivalence problem for the class of linear second order
hy\-per\-bo\-lic equations in two independent variables under an action of the
pseudo-group of con\-tact  transformations. {\'E}. Cartan's method is used for 
finding the Maurer -- Cartan forms for symmetry groups of equations from
the class and com\-pu\-ting structure equations and complete sets of
differential invariants for these groups.  The solution of the equivalence
problem is formulated in terms of these dif\-fe\-ren\-ti\-al invariants.
\end{abstract}


\section*{Introduction}

In the present paper, we find neccessary and sufficient conditions for two
equations from the class of linear second order hyperbolic equations
\begin{equation}
u_{tx} = T(t,x)\,u_t+X(t,x)\,u_x+U(t,x)\,u
\label{hyper}
\end{equation}
\noindent
to be equivalent under an action of the contact transformation pseudo-group. We use \'Elie
Cartan's method of equivalence, \cite{Cartan1} - \cite{Cartan5}, in its form
developed by Fels and Olver, \cite{FO,FO2}, to com\-pute the Maurer -- Cartan
forms,  the structure equations, the basic invariants, and the invariant derivatives  for symmetry 
groups of equations from the class . All differential invariants are functions of the basic invariants 
and their invariant derivatives. The differential invariants parametrize classifying manifolds associated with given equations. 
Cartan's solution to the equivalence problem states that two equations are (locally) 
equivalent if and only if their classifying manifolds (locally) over\-lap.

The sym\-met\-ry clas\-si\-fi\-ca\-tion problem for classes of
dif\-fe\-ren\-ti\-al equa\-ti\-ons is closely related to the problem of local
equivalence: symmetry groups of two equations are necessarily isomorphic if these equations 
are equi\-va\-lent, while the converse statement is not true in general.
The symmetry analysis of linear second order hyperbolic equations (\ref{hyper})
is done by Lie, \cite[Vol.~3, pp~492-523]{Lie}.
Two semi-invariants $H = - T_t+T\,X+U$ and $K=-X_x+T\,X+U$ were discovered
by Laplace, \cite{Laplace}. These functions are unaltered under an action of
the pseudo-groups of linear transformations $\overline{u} = c(t,x)\,u$.
In \cite{Ovsiannikov1960}, Ovsiannikov found  the invariants $P=K\,H^{-1}$,
$Q=(\ln \vert\,H\,\vert)_{tx}\,H^{-1}$ and used them to classify equations
(\ref{hyper}) with non-trivial symmetry groups.
In \cite[th~2.3]{Ibragimov97}, \cite[\S~10.4.2]{Ibragimov99}, it was claimed
that the invariants $P$ and $Q$ form a basis of differential invariants for
equations (\ref{hyper}), while all the other invariants are functions of $P$
and $Q$ and their invariant derivatives.
In \cite{JMWS}, a basis of five invariants and operators of invariant
differentiation are found in the case $P_x \not = 0$.
In the case $P_t \not =0$ and $P_x \not = 0$ two bases of four invariants are computed in
\cite{Ibragimov2004}.

In \cite{LisleReid99}, the invariant version of Lie's in\-fi\-ni\-te\-si\-mal method was 
developed and applied to the symmetry classification of the class (\ref{hyper}).

The symmetry classification problem and invariants for the class of linear
parabolic equations $u_{xx} = T(t,x)\,u_t+X(t,x)\,u_x+U(t,x)\,u$ are studied
in \cite{Lie,Ovsiannikov,Ibragimov2002,JM} by Lie's infinitesimal method.
In \cite{Morozov03,Morozov04}, Cartan's method is applied to solve the contact
equi\-va\-len\-ce problem for this class.

The paper is organized as follows. In Section 1, we begin with some notation,
and use Cartan's equivalence method to find the invariant 1-forms and the
structure equa\-ti\-ons for the pseudo-group of contact transformations on the
bundle of second-order jets. In Section 2, we briefly describe the approach to
computing Maurer - Cartan forms and structure equations for symmetry groups of
differential equations via the moving co\-fra\-me method of Fels and Olver. In
Section 3, the method is applied to the class of hy\-per\-bo\-lic equations
(\ref{hyper}). Finally, we make some concluding remarks.

\section{Pseudo-group of contact transformations}

In this paper, all considerations are of local nature, and all mappings are
real analytic. Let $\Ec = \R^n \times \R \rightarrow \R^n$ be a trivial
bundle with the local base coordinates $(x^1,...,x^n)$ and the local fibre
coordinate $u$; then by $J^2(\Ec)$ denote the bundle of the second-order jets
of sections of $\Ec$, with the local coordinates $(x^i,u,p_i,p_{ij})$,
$i,j\in\{1,...,n\}$, $i \le j$. For every local section $(x^i,f(x))$ of $\Ec$,
the corresponding 2-jet $(x^i,f(x),\partial f(x)/\partial x^i,
\partial^2 f(x)/\partial x^i\partial x^j)$ is denoted by $j_2(f)$.
A differential 1-form $\vartheta$ on $J^2(\Ec)$ is called a {\it contact
form} if it is annihilated by all 2-jets of local sections:
$j_2(f)^{*}\vartheta = 0$. In the local coordinates every contact 1-form is a
linear combination of the forms  $\vartheta_0 = du - p_{i}\,dx^i$,
$\vartheta_i = dp_i - p_{ij}\,dx^j$, $i, j \in \{1,...,n\}$, $p_{ji} = p_{ij}$
(here and later we use the Einstein summation convention, so
$p_i\,dx^i = \sum_{i=1}^{n}\,p_i\,dx^i$, etc.)
A local diffeomorphism
\begin{equation}
\Delta : J^2(\Ec) \rightarrow J^2(\Ec),
\qquad
\Delta : (x^i,u,p_i,p_{ij}) \mapsto
(\overline{x}^i,\overline{u},\overline{p}_i,\overline{p}_{ij}),
\label{Delta}
\end{equation}
\noindent
is called a {\it contact transformation} if for every contact 1-form
$\vartheta$ the form $\Delta^{*}\overline{\vartheta}$ is also contact.
We use Cartan's method of equivalence, \cite{Cartan5,Olver95}, to obtain a collection of 
in\-va\-ri\-ant 1-forms for the pseudo-group of contact transformations on $J^2(\Ec)$. For this, 
take the coframe $\{\vartheta_0, \vartheta_i, dx^i, dp_{ij}\,\vert\,i,j\in\{1,...,n\}, i\le j\}$
on $J^2(\Ec)$. A contact transformation (\ref{Delta}) acts on this coframe in
the following manner:
\[
\Delta^{*}\,
\left(
\begin{array}{c}
\overline{\vartheta}_0\\
\overline{\vartheta}_i\\
d\overline{x}^i\\
d\overline{p}_{ij}
\end{array}
\right)=
S\,\left(\begin{array}{c}
\vartheta_0\\
\vartheta_k\\
d x^k\\
d p_{kl}
\end{array}\right),
\]
\noindent
where $S : J^2(\Ec) \rightarrow \Gc $ is an analytic function, and $\Gc$ is the Lie
group of non-degenerate block matrices of the form
\[
\left(
\begin{array}{cccc}
a            & \wt{a}^k        & 0               & 0                 \\
\wt{g}_i     & h_{i}^{k}       & 0               & 0                 \\
\wt{c}^i     & \wt{f}^{ik}     & b_{k}^{i}       & r^{ikl}           \\
\wt{s}_{ij}  & \wt{w}_{ij}^{k} & \wt{z}_{ijk}    & \wt{q}_{ij}^{kl}
\end{array}
\right).
\]
\noindent
In these matrices,  $i,j,k,l \in \{1,...,n\}$, $r^{ikl}$ are defined for $k\le l$,
$\wt{s}_{ij}$, $\wt{w}_{ij}^{k}$,  and $\wt{z}_{ijk}$ are defined for
$i\le j$, and  $\wt{q}_{ij}^{kl}$ are defined for $i\le j$, $k\le l$.

Let us show that $\wt{a}^k = 0$. Indeed, the exterior (non-closed!) ideal
$\Ic = {\rm span} \{\vartheta_0, \vartheta_i\}$ has the derived ideal
$\delta \Ic = \{\omega\in \Ic \,\vert\, d\omega \in \Ic\}
= {\rm span}\{\vartheta_0\}$. Since $\Delta^{*}\,\overline{\Ic} \subset \Ic$
implies $\Delta^{*}\,(\delta\,\overline{\Ic}) \subset
\delta(\Delta^{*}\,\overline{\Ic}) \subset
\delta\,\Ic$, we obtain $\Delta^{*}\,\overline{\vartheta}_0 = a\,\vartheta_0$.

For convenience in the following computations, we denote by $(B_i^j)$ the
inverse matrix for $(b_i^j)$, so $b_i^j\,B_j^k = \delta_i^k$, by $(H_i^j)$
denote the inverse matrix for $(h_i^j)$, so $h_i^j\,H_j^k = \delta_i^k$,
change the variables on $\Gc$ such that
$g_i= \wt{g}_i\,a^{-1}$,
$f^{ij} = \wt{f}^{ik}\,H_{k}^{j}$,
$c^{i}=\wt{c}^{i}\,a^{-1} - f^{ik}\,g_k$,
$s_{ij} = \wt{s}_{ij}\,a^{-1} - \wt{w}_{ij}^{k}\,H_k^m\,g_m -
\wt{z}_{ijm}\,B_k^m\,c^k$,
$w_{ij}^k=\wt{w}_{ij}^m\,H_m^k - \wt{z}_{ijm}\,B_l^m\,f^{lk}$,
$z_{ijk} = \wt{z}_{ijm}\,B_k^m$,
$q_{ij}^{kl} = \wt{q}_{ij}^{kl}
- \wt{z}_{ijm}\,B_{m^{\prime}}^m\,r^{m^{\prime}kl}$,
and define $Q_{k^{\prime}l^{\prime}}^{kl}$ by
$Q_{k^{\prime}l^{\prime}}^{kl}\,q^{k^{\prime}l^{\prime}}_{ij} =
\delta_i^k\,\delta_j^l$.

In accordance with Cartan's method of equivalence, we take the lifted
coframe
\begin{equation}
\left(
\begin{array}{c}
\Theta_0\\
\Theta_i\\
\Xi^i\\
\Sigma_{ij}
\end{array}
\right)=
S\,
\left(\begin{array}{c}
\vartheta_0\\
\vartheta_k\\
dx^k\\
dp_{kl}
\end{array}\right)
=
\left(\begin{array}{l}
a\,\vartheta_0\\
g_i\,\Theta_0 + h_i^k\,\vartheta_k\\
c^i\,\Theta_0+f^{ik}\,\Theta_k+b_k^i\,dx^k + r^{ikl}\,dp_{kl}\\
s_{ij}\,\Theta_0+w_{ij}^{k}\,\Theta_k+z_{ijk}\,\Xi^k + q_{ij}^{kl}\,dp_{kl}
\end{array}\right)
\label{LCF}
\end{equation}
\noindent
on $J^2(\Ec)\times\Gc$. Expressing $du$, $dx^k$, $d p_k$, and $dp_{kl}$ from
(\ref{LCF}) and substituting them to  $d\Theta_0$, we have
\[\fl
d\Theta_0 = da\w \vartheta_0 + a\,d\vartheta_0
= da\,a^{-1}\w \Theta_0 + a\,dx^i\w dp_i
= da\,a^{-1}\w \Theta_0 + a\,dx^i\w \vartheta_i
\]
\[\fl\hspace{25pt}
= \Phi_0^0 \w\Theta_0 + a\,B^i_k\,H^m_i \,\Xi^k \w \Theta_m
+  a\,H^m_i\,R^{ikl}\,\Sigma_{kl}\w\Theta_m
\]
\begin{equation}
\fl\hspace{25pt}
+ a\,H^m_i\,\left(B^i_k\,f^{kj}
+R^{ikl}\,w^j_{kl}\right)\,\Theta_j\w\Theta_m,
\label{dTheta0}
\end{equation}
\noindent
where
\[\fl
\Phi^0_0 = da\,a^{-1} + a\,H^{m^{\prime}}_i\,\left(
B^i_k\left(c^k+R^{ikl}\,s_{kl}\right)\,\Theta_{m^{\prime}}
-g_{m^{\prime}}\,B_k^i\,(\Xi^k-c^k\,\Theta_0-f^{kj}\,\Theta_j)
     \right.
     \]
     \[\fl\hspace{25pt}
     \left.
-g_{m^{\prime}}\,R^{ikl}\,(\Sigma_{kl} - s_{kl}\,\Theta_0-w_{kl}^m\,\Theta_m
-z_{klm}\,\Xi^m)
\right)
\]
\noindent
and
$R^{jkl} = - r^{ik^{\prime}l^{\prime}}\,B_i^j\,Q_{k^{\prime}l^{\prime}}^{kl}$.

The multipliers of $\Xi^k \w \Theta_m$, $\Sigma_{kl}\w\Theta_m$, and
$\Theta_j\w\Theta_m$ in (\ref{dTheta0}) are essential  torsion coefficients.
We normalize them by setting $a\,B^i_k\,H^m_i = \delta^m_k$, $R^{ikl} = 0$,
and $f^{kj} = f^{jk}$. Therefore the first normalization is
\begin{equation}
h^k_i = a\,B^k_i, \qquad r^{ikl} = 0, \qquad  f^{kj} = f^{jk}.
\label{normalize1}
\end{equation}
\noindent

Analysing $d\Theta_i$, $d\Xi^i$, and $d\Sigma_{ij}$ in the same way,
we obtain the following nor\-ma\-li\-za\-ti\-ons:
\begin{equation}
q^{kl}_{ij} = a\,B^k_i\,B^l_j, \qquad s_{ij} = s_{ji}, \qquad
w^k_{ij} = w^k_{ji},\qquad z_{ijk} = z_{jik} = z_{ikj}.
\label{normalize2}
\end{equation}
\noindent
After these reductions  the structure equations for the lifted coframe
have the form
\[\fl
d \Theta_0 = \Phi^0_0 \w \Theta_0 + \Xi^i \w \Theta_i,
\]
\[\fl
d \Theta_i = \Phi^0_i \w \Theta_0 + \Phi^k_i \w \Theta_k
+ \Xi^k \w \Sigma_{ik},
\]
\[\fl
d \Xi^i = \Phi^0_0 \w \Xi^i -\Phi^i_k \w \Xi^k
+\Psi^{i0} \w \Theta_0
+\Psi^{ik} \w \Theta_k,
\]
\[\fl
d \Sigma_{ij} = \Phi^k_i \w \Sigma_{kj} - \Phi^0_0 \w \Sigma_{ij}
+ \Upsilon^0_{ij} \w \Theta_0
+ \Upsilon^k_{ij} \w \Theta_k + \Lambda_{ijk} \w \Xi^k,
\]
\noindent
where the forms $\Phi^0_0$, $\Phi^0_i$, $\Phi^k_i$, $\Psi^{i0}$, $\Psi^{ij}$,
$\Upsilon^0_{ij}$, $\Upsilon^k_{ij}$, and $\Lambda_{ijk}$ are defined by the
following equations:
\[\fl
\Phi^0_0 = da\,a^{-1} - g_k\,\Xi^k+(c^k+f^{km}\,g_m)\,\Theta_k,
\]
\[\fl
\Phi^0_i = d g_i+g_k\,db^k_j\,B^j_i-(g_i\,g_k+s_{ik}+c^j\,z_{ijk})\,\Xi^k
+c^k\,\Sigma_{ik}
\]
\[
+(g_i\,c^k+g_i\,g_m\,f^{mk}-c^j\,w^k_{ij}+f^{mk}\,s_{im})\,\Theta_k,
\]
\[\fl
\Phi^k_i=\delta^k_i\,da\,a^{-1}-d b^k_j\,B^j_i
+(g_i\,\delta^k_j-w^k_{ij}-f^{km}\,z^i_{jm})\,\Xi^j+f^{km}\,\Sigma_{im}
+f^{jm}\,w^k_{ij}\,\Theta_m,
\]
\[\fl
\Psi^{i0} = dc^i+f^{ij}\,\Phi^0_j+c^k\,\Phi^i_k
+(c^i\,f^{mj}\,g_m-c^k\,f^{mj}\,w^i_{kj})\,\Theta_j-c^k\,f^{ij}\,\Sigma_{kj}
\]
\begin{equation}
+c^k\,(f^{im}\,z^{kmj}+w^i_{kj}-g_k\,\delta^i_j-g_j\,\delta^i_k)\,\Xi^j,
\label{PCF_contact}
\end{equation}
\[\fl
\Psi^{ij} = df^{ij}+(f^{ik}\,\delta^j_m+f^{jk}\,\delta^i_m)\,\Phi^m_k
+(c^i\,\delta^j_k+c^j\,\delta^i_k-f^{ij}\,g_k+f^{im}\,f^{jl}\,z_{klm})\,\Xi^k
\]
\[
+f^{ij}\,(c^k+f^{km}\,g_m)\,\Theta_k - f^{ik}\,f^{jm}\,\Sigma_{km},
\]
\[\fl
\Upsilon^0_{ij} = d s_{ij}-s_{ij}\,da\,a^{-1}+s_{kj}\,db^k_m\,B^m_i
+s_{ik}\,db^k_m\,B^m_j+s_{ij}\,\Phi^0_0+w^k_{ij}\,\Phi^0_k+z_{ijk}\,\Psi^{k0},
\]
\[\fl
\Upsilon^k_{ij} = dw^k_{ij} - w^k_{ij}\,da\,a^{-1}
+(w^k_{il}\,\delta^{m^{\prime}}_j+w^k_{jl}\,\delta^{m^{\prime}}_i)\,db^l_m\,
B^m_{m^{\prime}}
+(s_{ij}\,\delta^k_m+z_{ijl}\,f^{m^{\prime}k}\,w^l_{m^{\prime}m})\,\Xi^m
\]
\[
+w^m_{ij}\,\Phi^k_m
+f^{lk}\,(w^m_{il}\,\delta^{m^{\prime}}_j
+w^m_{jl}\,\delta^{m^{\prime}}_i)\,\Sigma_{m^{\prime}m}
-(c^k+f^{mk}\,g_m)\,\Sigma_{ij},
\]
\[\fl
\Lambda_{ijk} = d z_{ijk} - 2\,z_{ijk}\,da\,a^{-1}
+ z_{ijl}\,db^l_m\,B^m_k
+ z_{ilk}\,db^l_m\,B^m_j
+ z_{ljk}\,db^l_m\,B^m_i
+z_{ijk}\,\Phi^0_0
\]
\[
+z_{ijk}\,g_m\,\Xi^m
+g_i\,\Sigma_{jk}
+g_j\,\Sigma_{ik}
+g_k\,\Sigma_{ij}
-w^l_{ij}\,\Sigma_{lk}
-w^l_{ik}\,\Sigma_{lj}
-w^l_{jk}\,\Sigma_{li}
\]
\[
-f^{lm}\,(
z_{imj}\,\Sigma_{kl}
+z_{imk}\,\Sigma_{jl}
+z_{jmk}\,\Sigma_{il}
).
\]

Let $\Hc$ be the subgroup of $\Gc$ defined by (\ref{normalize1}) and
(\ref{normalize2}). We shall prove that the restriction of the lifted coframe
(\ref{LCF}) to $J^2(\Ec) \times \Hc$ satisfies  Cartan's test of involutivity,
\cite[def~11.7]{Olver95}. The structure equations remain unchanged under the
following transformation of the forms
(\ref{PCF_contact}):
$\Phi^0_0 \mapsto \wt{\Phi}^0_0$,
$\Phi^k_i \mapsto \wt{\Phi}^k_i$,
$\Phi^0_i \mapsto \wt{\Phi}^0_i$,
$\Psi^{ij} \mapsto \wt{\Psi}^{ij}$,
$\Psi^{i0} \mapsto \wt{\Psi}^{i0}$,
$\Upsilon^{0}_{ij} \mapsto \wt{\Upsilon}^{0}_{ij}$,
$\Upsilon^{k}_{ij} \mapsto \wt{\Upsilon}^{k}_{ij}$,
$\Lambda_{ijk} \mapsto \wt{\Lambda}_{ijk}$,
where
\[
\wt{\Phi}^0_0 = \Phi^0_0 + K\,\Theta_0,
\]
\[
\wt{\Phi}^k_i = \Phi^k_i + L^{kl}_i\,\Theta_l + M^k_i\,\Theta_0,
\]
\[
\wt{\Phi}^0_i = \Phi^0_i + M^k_i\,\Theta_k + N_i\,\Theta_0,
\]
\[
\wt{\Psi}^{ij} = \Psi^{ij} + P^{ij}\,\Theta_0+S^{ijk}\,\Theta_k
- L^{ij}_k\,\Xi^k,
\]
\begin{equation}
\wt{\Psi}^{i0} = \Psi^{i0} + P^{ij}\,\Theta_j +T^i\,\Theta_0 + K\,\Xi^i
- M^i_k\,\Xi^k,
\label{PCF_transform}
\end{equation}
\[
\wt{\Upsilon}^{0}_{ij} = \Upsilon^{0}_{ij} + U_{ij}\,\Theta_0
+ V^k_{ij}\,\Theta_k + W_{ijk}\,\Xi^k + K\,\Sigma_{ij} + M^k_i\,\Sigma_{kj},
\]
\[
\wt{\Upsilon}^{k}_{ij} = \Upsilon^{k}_{ij} + X^{kl}_{ij}\,\Theta_l
+ V^k_{ij}\,\Theta_0 + Y^{k}_{ijl}\,\Xi^l + L_i\,\Sigma_{lj},
\]
\[
\wt{\Lambda}_{ijk} = \Lambda_{ijk} + Z_{ijkl}\,\Xi^l + Y^l_{ijk}\,\Theta_l
+ W_{ijk}\,\Theta_0,
\]
\noindent
and $K$, $L^{kl}_i$, $M^k_i$, $N_i$, $P^{ij}$, $S^{ijk}$, $T^i$, $U_{ij}$,
$V^k_{ij}$, $W_{ijk}$, $X^{kl}_{ij}$, $Y^k_{ijl}$, and $Z_{ijkl}$ are
arbitrary constants satisfying the following symmetry conditions :
\[
L^{kl}_i=L^{lk}_i,
\quad
P^{ij}=P^{ji},
\quad
S^{ijk}=S^{jik}=S^{ikj},
\quad
U_{ij}=U_{ji},
\quad
V^k_{ij}=V^k_{ji},
\]
\begin{equation}
W_{ijk}=W_{jik}=W_{ikj},
\quad
X^{kl}_{ij}=X^{kl}_{ji}=X^{lk}_{ij},
\quad
Y^k_{ijl}=Y^k_{jil}=Y^k_{ilj},
\label{symm_cond}
\end{equation}
\[
Z_{ijkl}=Z_{jikl}=Z_{ijlk}=Z_{ikjl}.
\]
\noindent
The number of such constants
\[\fl
r^{(1)} = 1 + {{n^2\,(n+1)}\over{2}} +n^2 + n + {{n\,(n+1)}\over{2}}
+ {{n\,(n+1)\,(n+2)}\over{6}}
+ n
+ {{n\,(n+1)}\over{2}}
\]
\[\fl
\hspace{25pt}
+ {{n^2\,(n+1)}\over{2}}
+ {{n\,(n+1)\,(n+2)}\over{6}}
+ {{n^2\,(n+1)^2}\over{4}}
+ {{n^2\,(n+1)\,(n+2)}\over{6}}
\]
\[\fl
\hspace{25pt}
+ {{n\,(n+1)\,(n+2)\,(n+3)}\over{24}} =
{{1}\over{24}}\,(n+1)\,(n+2)\,(11\,n^2+29\,n+12)
\]
\noindent
is the degree of indeterminancy  of the lifted coframe,
\cite[def~11.2]{Olver95}. The reduced cha\-rac\-ters of this coframe,
\cite[def~11.4]{Olver95},  are easily found:
$s^{\prime}_i = {{1}\over{2}}\,(n+1)\,(n+4) - i$
when $i \in \{1, ..., n+1\}$ and 
$s^{\prime}_{n+1+j} = {{1}\over{2}}\,(n+1-j)\,(n+2-j)$ when $j \in \{1, ..., n\}$.
A simple cal\-cu\-la\-ti\-on shows that
$r^{(1)} =
s^{\prime}_1
+ 2\,s^{\prime}_2
+ 3\,s^{\prime}_3
+ ...
+ (2\,n+1)\,s^{\prime}_{2\,n+1}$.
So the Cartan test is satisfied, and the lifted coframe is involutive.

It is easy to directly verify that a transformation
$\widehat{\Delta} : J^2(\Ec) \times \Hc \rightarrow J^2(\Ec) \times \Hc$
satisfies the conditions
\begin{equation}
\widehat{\Delta}^{*}\, \overline{\Theta}_0 = \Theta_0,
\qquad
\widehat{\Delta}^{*}\, \overline{\Theta}_i = \Theta_i,
\qquad
\widehat{\Delta}^{*}\, \overline{\Xi}^i = \Xi^i,
\qquad
\widehat{\Delta}^{*}\, \overline{\Sigma}_{ij} = \Sigma_{ij}
\label{def_cond}
\end{equation}
\noindent
if and only if it is projectable on $J^2(\Ec)$, and its projection
$\Delta : J^2(\Ec) \rightarrow J^2(\Ec)$ is a contact transformation.

Since (\ref{def_cond}) imply
$\widehat{\Delta}^{*}\,d\overline{\Theta}_0 = d\Theta_0$,
$\widehat{\Delta}^{*}\,d\overline{\Theta}_i = d\Theta_i$,
$\widehat{\Delta}^{*}\,d\overline{\Xi}^i = d\Xi^i$,
and
$\widehat{\Delta}^{*}\,d\overline{\Sigma}_{ij} = d\Sigma_{ij}$,
we have
\[\fl
\widehat{\Delta}^{*} \left(\overline{\Phi}^0_0 \w \overline{\Theta}_0
+ \overline{\Xi}^i \w \overline{\Theta}_i\right)
=
\left(\widehat{\Delta}^{*} \overline{\Phi}^0_0\right) \w \Theta_0
+ \Xi^i \w \Theta_i
=
\Phi^0_0 \w \Theta_0 + \Xi^i \w \Theta_i,
\]
\[\fl
\widehat{\Delta}^{*} \left(\overline{\Phi}^0_i \w \overline{\Theta}_0
+ \overline{\Phi}^k_i \w \overline{\Theta}_k
+ \overline{\Xi}^k \w \overline{\Sigma}_{ik}\right)
=
\widehat{\Delta}^{*} \left(\overline{\Phi}^0_i\right) \w \Theta_0
+ \widehat{\Delta}^{*} \left(\overline{\Phi}^k_i\right) \w \Theta_k
+ \Xi^k \w \Sigma_{ik}
\]
\[\fl\hspace{25pt}
=
\Phi^0_i \w \Theta_0
+ \Phi^k_i \w \Theta_k
+ \Xi^k \w \Sigma_{ik},
\]
\[\fl
\widehat{\Delta}^{*}\left(\overline{\Phi}^0_0 \w \overline{\Xi}^i
- \overline{\Phi}^i_k \w \overline{\Xi}^k
+\overline{\Psi}^{i0} \w \overline{\Theta}_0
+\overline{\Psi}^{ik} \w \overline{\Theta}_k\right)
\]
\[\fl\hspace{25pt}
=
\widehat{\Delta}^{*}\left(\overline{\Phi}^0_0\right) \w \Xi^i
-\widehat{\Delta}^{*}\left(\overline{\Phi}^i_k\right) \w \Xi^k
+\widehat{\Delta}^{*}\left(\overline{\Psi}^{i0}\right) \w \Theta_0
+\widehat{\Delta}^{*}\left(\overline{\Psi}^{ik}\right) \w \Theta_k
\]
\[\fl\hspace{25pt}
=
\Phi^0_0 \w \Xi^i -\Phi^i_k \w \Xi^k
+\Psi^{i0} \w \Theta_0
+\Psi^{ik} \w \Theta_k,
\]
\[\fl
\widehat{\Delta}^{*}\left(
\overline{\Phi}^k_i \w \overline{\Sigma}_{kj}
- \overline{\Phi}^0_0 \w \overline{\Sigma}_{ij}
+ \overline{\Upsilon}^0_{ij} \w \overline{\Theta}_0
+ \overline{\Upsilon}^k_{ij} \w \overline{\Theta}_k
+ \overline{\Lambda}_{ijk} \w \overline{\Xi}^k
\right)
\]
\[\fl\hspace{25pt}
=
\widehat{\Delta}^{*}\left(\overline{\Phi}^k_i\right) \w \Sigma_{kj}
- \widehat{\Delta}^{*}\left(\overline{\Phi}^0_0\right) \w \Sigma_{ij}
+ \widehat{\Delta}^{*}\left(\overline{\Upsilon}^0_{ij}\right) \w \Theta_0
+ \widehat{\Delta}^{*}\left(\overline{\Upsilon}^k_{ij}\right) \w \Theta_k
\]
\[\fl\hspace{25pt}
+ \widehat{\Delta}^{*}\left(\overline{\Lambda}_{ijk}\right) \w \Xi^k
=
\Phi^k_i \w \Sigma_{kj}
- \Phi^0_0 \w \Sigma_{ij}
+ \Upsilon^0_{ij} \w \Theta_0
+ \Upsilon^k_{ij} \w \Theta_k
+ \Lambda_{ijk} \w \Xi^k.
\]
\noindent
Therefore, we have the following transformation rules
\[
\widehat{\Delta}^{*}\left(\overline{\Phi}^0_0\right) = \wt{\Phi}^0_0,
\quad
\widehat{\Delta}^{*}\left(\overline{\Phi}^k_i\right) = \wt{\Phi}^k_i,
\quad
\widehat{\Delta}^{*}\left(\overline{\Phi}^0_i\right) = \wt{\Phi}^0_i,
\]
\begin{equation}
\widehat{\Delta}^{*}\left(\overline{\Psi}^{ij}\right) = \wt{\Psi}^{ij},
\quad
\widehat{\Delta}^{*}\left(\overline{\Psi}^{i0}\right) = \wt{\Psi}^{i0},
\quad
\widehat{\Delta}^{*}\left(\overline{\Upsilon}^{0}_{ij}\right) =
\wt{\Upsilon}^{0}_{ij},
\label{Gamma_rules}
\end{equation}
\[
\widehat{\Delta}^{*}\left(\overline{\Upsilon}^{k}_{ij}\right) =
\wt{\Upsilon}^{k}_{ij},
\quad
\widehat{\Delta}^{*}\left(\overline{\Lambda}_{ijk}\right) = \wt{\Lambda}_{ijk}.
\]
\noindent
where the constants $K$, ..., $Z_{ijkl}$ in (\ref{PCF_transform}) are replaced
by arbitrary functions on $J^2(\Ec) \times \Hc$ such that the same symmetry
conditions (\ref{symm_cond}) are satisfied.

\section{Contact symmetries of differential equations}

Suppose $\Rc$ is a second-order differential equation in one dependent and
$n$ independent variables. We consider $\Rc$ as a sub-bundle in $J^2(\Ec)$.
Let $Cont(\Rc)$ be the group of contact symmetries for $\Rc$. It consists of
all the contact transformations on $J^2(\Ec)$ mapping $\Rc$ to itself.
The moving coframe method, \cite{FO,FO2}, is applicable to find invariant
1-forms characterizing $Cont(\Rc)$ is the same way, as the restriction of
the lifted coframe (\ref{LCF}) to $J^2(\Ec)\times\Hc$ characterizes
$Cont(J^2(\Ec))$. We briefly outline this approach.

Let $\iota : \Rc \rightarrow J^2(\Ec)$ be an embedding. The invariant
1-forms of $Cont(\Rc)$ are re\-stric\-ti\-ons of the coframe (\ref{LCF}),
(\ref{normalize1}), (\ref{normalize2}) to $\Rc$:
$\theta_0 = \iota^{*} \Theta_0$, $\theta_i= \iota^{*}\Theta_i$,
$\xi^i = \iota^{*}\Xi^i$, and $\sigma_{ij}=\iota^{*}\Sigma_{ij}$
(for brevity we identify the map
$\iota \times id : \Rc\times \Hc \rightarrow J^2(\Ec)\times \Hc$ with
$\iota : \Rc \rightarrow J^2(\Ec)$). The forms $\theta_0$, $\theta_i$,
$\xi^i$, and $\sigma_{ij}$ have some linear dependencies, i.e., there exists a
non-trivial set of functions $E^0$, $E^i$, $F_i$, and $G^{ij}$ on
$\Rc\times \Hc$ such that
$E^0\,\theta_0 + E^i\,\theta_i + F_i\,\xi^i+ G^{ij}\,\sigma_{ij} \equiv 0$.
These functions are lifted invariants of $Cont(\Rc)$. Setting them equal to
some constants allows us to specify some parameters $a$, $b^k_i$, $c_i$,
$g_i$, $f^{ij}$, $s_{ij}$, $w^k_{ij}$, and $z_{ijk}$ of the group $\Hc$
as functions of the coordinates on $\Rc$ and the other group parameters.

After these normalizations, a part of the forms $\phi^0_0=\iota^{*}\Phi^0_0$, 
$\phi^k_i=\iota^{*}\Phi^k_i$, $\phi^0_i=\iota^{*}\Phi^0_i$, 
$\psi^{ij}=\iota^{*}\Psi^{ij}$, $\psi^{i0}=\iota^{*}\Psi^{i0}$, 
$\upsilon^{0}_{ij}=\iota^{*}\Upsilon^{0}_{ij}$,
$\upsilon^{k}_{ij}=\iota^{*}\Upsilon^{k}_{ij}$, and
$\lambda_{ijk}=\iota^{*}\Lambda_{ijk}$, or some their li\-ne\-ar combinations,
become semi-basic, i.e., they do not include the differentials of the
pa\-ra\-me\-ters of $\Hc$. From (\ref{Gamma_rules}) and (\ref{PCF_transform}),
we have the following statements:
(i) if $\phi^0_0$ is semi-ba\-sic, then its coefficients at $\theta_k$, $\xi^k$,
and $\sigma_{kl}$ are lifted invariants of $Cont(\Rc)$;
(ii) if $\phi^0_i$ or $\phi^k_i$ are semi-basic, then their coefficients at
$\xi^k$ and $\sigma_{kl}$ are lifted invariants of $Cont(\Rc)$;
(iii) if $\psi^{i0}$, $\psi^{ij}$, or $\lambda_{ijk}$ are semi-basic,
then their coefficients  at $\sigma_{kl}$ are lifted invariants of $Cont(\Rc)$.
Setting these invariants equal to some constants, we get specifications of some
more parameters of $\Hc$ as functions of the coordinates on $\Rc$ and the other
group parameters.

More lifted invariants can appear as essential torsion coefficients in the
reduced structure equations
\[
d \theta_0 = \phi^0_0 \w \theta_0 + \xi^i \w \theta_i
\]
\[
d \theta_i =
\phi^0_i \w \theta_0 + \phi^k_i \w \theta_k + \xi^k \w \sigma_{ik}
\]
\[
d \xi^i = \phi^0_0 \w \xi^i -\phi^i_k \w \xi^k +\psi^{i0} \w \theta_0
+\psi^{ik} \w \theta_k
\]
\[
d \sigma_{ij} = \phi^k_i \w \sigma_{kj} - \phi^0_0 \w \sigma_{ij}
+ \upsilon^0_{ij} \w \theta_0 + \upsilon^k_{ij} \w \theta_k
+ \lambda_{ijk} \w \xi^k.
\]
\noindent
After normalizing these invariants and repeating the process, two outputs are
possible. In the first case, the reduced lifted coframe appears to be
involutive. Then this coframe is the desired set of defining forms for
$Cont(\Rc)$. In the second case, when the reduced lifted coframe does not
satisfy Cartan's test, we should use the procedure of prolongation,
\cite[ch~12]{Olver95}.

\section{Structure and invariants of symmetry groups for linear hyperbolic
equations}

We apply the method described in the previous section to the class of linear
hyperbolic equations (\ref{hyper}). Denote $x^1=t$, $x^2=x$, $p_1=u_t$,
$p_2=u_x$, $p_{11}=u_{tt}$, $p_{12}=u_{tx}$, and $p_{22}=u_{xx}$.
The coordinates on $\Rc$ are $\{(t, x, u, u_t, u_x, u_{tt}, u_{xx})\}$,
and the embedding $\iota : \Rc \rightarrow J^2(\Ec)$ is defined by
(\ref{hyper}). At the first step, we analyse the linear dependence between the
reduced forms $\theta_0$, $\theta_i$, $\xi^i$, and $\sigma_{ij}$. Without loss
of generality we suppose that $b^1_1 \not = 0$ and $b^2_2 \not = 0$, then we
find
$\sigma_{12} = E_1\,\sigma_{11}+E_2\,\sigma_{22}
+ E_3\,\theta_0
+ E_4\,\theta_1
+ E_5\,\theta_2
+ E_6\,\xi^1
+ E_7\,\xi^2$,
where, for example,
$E_1 = -(b^1_1\,b^2_2+b^1_2\,b^2_1)^{-1}\,b^1_1\,b^1_2$
and
$E_2 = -(b^1_1\,b^2_2+b^1_2\,b^2_1)^{-1}\,b^2_2\,b^2_1$.
Setting $E_1$, $E_2$, ..., $E_7$ equal to 0 sequentially, we have
$E_1 = 0 \Rightarrow b^1_2 = 0$,
$E_2 = 0 \Rightarrow b^2_1 = 0$,
$E_3 = 0 \Rightarrow
s_{12} =-z_{112}\,c^1-z_{122}\,c^2+ g_1\,(b^2_2)^{-1}\,T
+g_2\,(b^1_1)^{-1}X -(b^1_1\,b^2_2)^{-1}\,U
$,
$E_4 = 0 \Rightarrow
w^1_{12}=-z_{112}\,f^{11}-z_{122}\,f^{12}-(b^2_2)^{-1}\,T$,
$E_5 = 0 \Rightarrow w^2_{12}=-z_{112}\,f^{12}-z_{122}\,f^{22}-(b^1_1)^{-1}\,X$,
$E_6 = 0 \Rightarrow z_{112}= -a\,(b^1_1)^{-2}(b^2_2)^{-1}\,
(T\,u_{tt}+(2\,T\,X+2\,U-H)\,u_t+(X_t+X^2)\,u_x+(U_t+X\,U)\,u)$,
and
$E_7 = 0 \Rightarrow
z_{122}=-a\,(b^1_1)^{-1}(b^2_2)^{-2}\,
(X\,u_{xx}+(T_x+T^2)\,u_t+(2\,T\,X+2\,U-K)\,u_x+(U_x+T\,U)\,u)$,
where
$H = - T_t+T\,X+U$ and $K=-X_x+T\,X+U$ are the Laplace invariants,
\cite{Laplace}, \cite[\S~9]{Ovsiannikov}.

At the second step, we analyse the semi-basic forms $\phi^i_j$ and $\phi^0_j$.
We have
\[
\phi^2_1 \equiv
f^{12}\,\sigma_{11}
+\left( g_1+(b^1_1)^{-1}X \right)\,\xi^2
\quad
(\rm{mod}\,\,
\theta_0,
\theta_1,
\theta_2,
\xi^1),
\]
\noindent
therefore we take $f^{12} = 0$, $g_1 = -(b^1_1)^{-1}\,X$. This yields
\[\fl
\phi^2_1 \equiv
\left(-w^2_{11}+a\,f^{22}\,(b^1_1)^{-2}(b^2_2)^{-1}\,
(
T\,u_{tt}+(2\,T\,X+2\,U-H)\,u_t+(X_t+X^2)\,u_x
       \right.
       \]
       \[\fl\hspace{15pt}
       \left.
+(U_t+X\,U)\,u
)
\right)\,\xi^1
\quad
(\rm{mod}\,\, \theta_0, \theta_1, \theta_2),
\]
\noindent
therefore we set
$w^2_{11}=a\,f^{22}\,(b^1_1)^{-2}(b^2_2)^{-1}\,
(
T\,u_{tt}+(2\,T\,X+2\,U-H)\,u_t+(X_t+X^2)\,u_x+(U_t+X\,U)\,u)$.

After that, we have
\[\fl
\phi^1_2 \equiv
\left(
g_2+(b^2_2)^{-1}\,T
\right)\,\xi^1
+\left(
-w^1_{22}+a\,f^{11}\,(b^1_1)^{-1}(b^2_2)^{-2}\,(
X\,u_{xx}+(T_x+T^2)\,u_t
       \right.
       \]
       \[\fl\hspace{15pt}
       \left.
+(2\,T\,X+2\,U-K)\,u_x+(U_x+T\,U)\,u
)
\right)\,\xi^2
\quad
(\rm{mod}\,\, \theta_0, \theta_1, \theta_2),
\]
so we set
$g_2 = - (b^2_2)^{-1}\,T$ and
$
w^1_{22}=a\,f^{11}\,(b^1_1)^{-1}(b^2_2)^{-2}\,(
X\,u_{xx}+(T_x+T^2)\,u_t+(2\,T\,X+2\,U-K)\,u_x+(U_x+T\,U)\,u
)
$.

Then we have
$\phi^0_1 \equiv c^1\,\sigma_{11} \,\,\,
(\rm{mod}\,\, \theta_0, \theta_1, \theta_2, \xi^1, \xi^2)$,
$\phi^0_2 \equiv c^2\,\sigma_{22} \,\,\,
(\rm{mod}\,\, \theta_0, \theta_1, \theta_2, \xi^1, \xi^2)$,
so we set $c^1=0$ and $c^2 = 0$. Now we obtain
\[
\phi^0_1 \equiv K \,(b^1_1)^{-1}(b^2_2)^{-1}\,\xi^2
\quad (\rm{mod}\,\, \theta_0, \theta_1, \theta_2),
\]
\begin{equation}
\phi^0_2 \equiv H \,(b^1_1)^{-1}(b^2_2)^{-1}\,\xi^1
\quad (\rm{mod}\,\, \theta_0, \theta_1, \theta_2).
\label{phi_0_2_H}
\end{equation}

There are two possibilities now: $H \equiv K \equiv 0$
or at least one of the Laplace invariants is not identically equal 0.

We denote by $\mathcal{S}_1$ the subclass of equations (\ref{hyper}) such that
$H \equiv K \equiv 0$. For an equation from $\mathcal{S}_1$ we use the
procedures of absorption and prolongation, \cite{Olver95}, to compute the
structure equations:
\[
d\theta_0=\eta_1\w\theta_0
+\xi^1\w\theta_1
+\xi^2\w\theta_2,
\]
\[
d\theta_1=
\eta_2\w\theta_1
+\xi^1\w\sigma_{11},
\]
\[
d\theta_2=
\eta_3\w\theta_2
+\xi^2\w\sigma_{22},
\]
\[
d\xi^1=
(\eta_1-\eta_2)\w\xi^1
+\eta_4\w\theta_1,
\]
\[
d\xi^2=(\eta_1-\eta_3)\,\w\xi^2
+\eta_5\w\theta_2,
\]
\[
d\sigma_{11}=
(2\,\eta_2-\eta_1)\w\sigma_{11}
+\eta_6\w\theta_1
+\eta_7\w\xi^1,
\]
\[
d\sigma_{22}=
(2\,\eta_3-\eta_1)\w\sigma_{22}
+\eta_8\w\theta_2
+\eta_9\w\xi^2,
\]
\[
d\eta_1=0,
\]
\[
d\eta_2=
\pi_1\w\theta_1
+\eta_4\w\sigma_{11}
-\eta_6\w\xi^1,
\]
\[
d\eta_3=
\pi_2\w\theta_2
+\eta_5\w\sigma_{22}
-\eta_8\w\xi^2,
\]
\[
d\eta_4=
-\pi_1\w\xi^1
+\pi_3\w\theta_1
+(\eta_1-2\,\eta_2)\w\eta_4,
\]
\[
d\eta_5=
-\pi_2\w\xi^2
+\pi_4\w\theta_2
+(\eta_1-2\,\eta_3)\w\eta_5,
\]
\[
d\eta_6=
2\,\pi_1\w\sigma_{11}
+\pi_5\w\theta_1
+\pi_6\w\xi^1
+(\eta_2-\eta_1)\w\eta_6
-\eta_4\w\eta_7,
\]
\[
d\eta_7=
\pi_6\w\theta_1
+\pi_7\w\xi^1
-3\,\eta_6\w\sigma_{11}
+(3\,\eta_2-2\,\eta_1)\w\eta_7,
\]
\[
d\eta_8=
2\,\pi_2\w\sigma_{22}
+\pi_8\w\theta_2
+\pi_9\w\xi^2
+(\eta_3-\eta_1)\w\eta_8
-\eta_5\w\eta_9,
\]
\[
d\eta_9=
\pi_9\w\theta_2
+\pi_{10}\w\xi^2
-3\,\eta_8\w\sigma_{22}
+(3\,\eta_3-2\,\eta_1)\w\eta_9.
\]
\noindent
In these equations, the forms $\eta_1$, ... , $\eta_9$ on $J^2(\Ec) \times \Hc$
depend on differentials of the parameters of $\mathcal{H}$, while the forms
$\pi_1$, ..., $\pi_{10}$ depend on differentials of the prolongation variables.
From the structure equations it follows that Cartan's test for the lifted
coframe $\{\theta_0, \theta_1, \theta_2, \xi^1, \xi^2, \sigma_{11},
\sigma_{22}, \eta_1, \eta_2, ..., \eta_9\}$ is satisfied, therefore the coframe
is involutive.

The same calculations show that the symmetry group of the linear wave equation
$u_{tx} = 0$ has the same structure equations, but with a different
lifted coframe. All the essential torsion coefficients in the structure
equations are constants. Thus, applying Theorem 15.12 of \cite{Olver95}, we
obtain the well-known result, \cite[\S~9]{Ovsiannikov}: every equation from
$\mathcal{S}_1$ is contact equivalent to the wave equation.

\vskip 5 pt

Now we return to the case of $H \not \equiv 0 $ or $K \not \equiv 0 $.
Since we can replace $H$ and $K$ by renaming the independent variables
$t \mapsto x$, $x \mapsto t$, we put $H \not \equiv 0$ with\-out loss
of generality. Then we use (\ref{phi_0_2_H}) and take $b^2_2= H\,(b^1_1)^{-1}$.
After this, the form $\phi^1_1+\phi^2_2-2\,\phi^0_0$ becomes semi-basic. Since
$\phi^1_1+\phi^2_2-2\,\phi^0_0 \equiv f^{11}\,\sigma_{11}
+f^{22}\,\sigma_{22} \,\,\,(\rm{mod}\,\, \theta_1, \theta_2, \xi^1, \xi^2)$,
we take $f^{11} = 0$ and $f^{22} = 0 $. Then we have
$\phi^1_1+\phi^2_2-2\,\phi^0_0 \equiv
-\left(w^1_{11}+ H^{-1}(b^1_1)^{-1}(H_t+2\,X\,H)\right)\,\xi^1
-\left(w^2_{22} + H^{-2} b^1_1\,(H_x+2\,T\,H)\right)\,\xi^2
\,\,\,(\rm{mod}\,\, \theta_1, \theta_2)$, so we take
$w^1_{11} = - H^{-1}(b^1_1)^{-1}(H_t+2\,X\,H)$ and
$w^2_{22} = - H^{-2} b^1_1\,(H_x+2\,T\,H)$.

At the third step, we analyse the structure equations. After absorption of
torsion they have the  form
\[\fl
d\theta_0 = \eta_1 \w \theta_0 + \xi^1\w\theta_1 + \xi^2\w\theta_2,
\]
\[\fl
d\theta_1 = \eta_2\w\theta_1+\xi^1\w\sigma_{11} - P\,\theta_0\w\xi^2,
\]
\[\fl
d\theta_2 = (2\,\eta_1-\eta_2)\w\theta_2-\theta_0\w\xi^1+\xi^2\w\sigma_{22},
\]
\[\fl
d\xi^1 = (\eta_1-\eta_2)\w\xi^1,
\]
\begin{equation}
\fl
d\xi^2 = (\eta_2-\eta_1)\w\xi^2,
\label{SE_P}
\end{equation}
\[\fl
d\sigma_{11} = (2\,\eta_2-\eta_1)\w\sigma_{11}+\eta_3\w\xi^1
- P_t\,(b^1_1)^{-1}\,\theta_0\w\xi^2
+ (Q+1-2\,P)\, \theta_1\w\xi^2,
\]
\[\fl
d\sigma_{22} = (3\,\eta_1-2\,\eta_2)\w\sigma_{22}
+\eta_4\w\xi^2+(P+Q-2)\,\theta_2\w\xi^1,
\]
\noindent
where the functions
$P = K\,H^{-1}$ and
$Q= (H\,H_{tx}-H_t\,H_x)\,H^{-3} = (\ln \vert\, H\,\vert)_{tx}\,H^{-1}$ are
invariants of the symmetry group, and the 1-forms $\eta_1$, ..., $\eta_4$
depend on differentials of parameters of the group $\mathcal{H}$ (these forms
are not neccessary the same as in the case of an equation from
$\mathcal{S}_1$).

\vskip 5 pt

We denote by $\mathcal{S}_2$ the subclass of equations (\ref{hyper})
such that $P_t \not \equiv 0$. This subclass is not empty, since, for
example, the equation $u_{tx} = t^2x^2\,u_t+ u$ belongs to $\mathcal{S}_2$.
For an equation from $\mathcal{S}_2$ we can normalize $P_t\,(b^1_1)^{-1}$,
the only essential torsion coefficient in the structure equations (\ref{SE_P}),
to $1$ by setting $b^1_1 = P_t$. Then, after prolongation, we have the
involutive lifted coframe
$\bftheta = \{\theta_0, \theta_1, \theta_2, \xi^1,
\xi^2, \sigma_{11}, \sigma_{22}, \eta_1, \eta_2, \eta_3\}$
with the structure equations
\[\fl
d\theta_0=
\eta_1\w\theta_0
+\xi^1\w\theta_1
+\xi^2\w\theta_2,
\]
\[\fl
d\theta_1=
\eta_1\w\theta_1
-P\,\theta_0\w\xi^2
-J_2\,\theta_1\w\xi^1
-J_1\,\theta_1\w\xi^2
+\xi^1\w\sigma_{11},
\]
\[\fl
d\theta_2=
\eta_1\w\theta_2
-\theta_0\w\xi^1
+J_2\,\theta_2\w\xi^1
+J_1\,\theta_2\w\xi^2
+\xi^2\w\sigma_{22},
\]
\[\fl
d\xi^1=
J_1\,\xi^1\w\xi^2,
\]
\begin{equation}\fl
d\xi^2=
J_2\,\xi^1\w\xi^2,
\label{SE_Q_2}
\end{equation}
\[\fl
d\sigma_{11}=
\eta_1\w\sigma_{11}
+\eta_2\w\xi^1
-\theta_0\w\xi^2
+(Q+1-2\,P)\,\theta_1\w\xi^2
+2\,J_1\,\xi^2\w\sigma_{11},
\]
\[\fl
d\sigma_{22}=
\eta_1\w\sigma_{22}
+\eta_3\w\xi^2
+(P-2+Q)\,\theta_2\w\xi^1
-2\,J_2\,\xi^1\w\sigma_{22},
\]
\[\fl
d\eta_1=
(P-1)\,\xi^1\w\xi^2,
\]
\[\fl
d\eta_2=
\pi_1\w\xi^1
+\eta_1\w\eta_2
-3\,J_1\,\eta_2\w\xi^2
+J_2\,\theta_0\w\xi^2
+(4\,P\,J_2-2\,Q\,J_2-\D_1(Q)
       \]
       \[\fl\hspace{15pt}
-2\,J_2+3)\,\theta_1\w\xi^2
+(2\,J_1\,J_2+2-3\,P+3\,Q-2\,\D_2(J_2))\,\xi^2\w\sigma_{11},
\]
\[\fl
d\eta_3=
\pi_2\w\xi^2
+\eta_1\w\eta_3
+3\,J_2\,\eta_3\w\xi^1
+(2\,J_1\,(P+Q-2)-\D_2(Q)-\D_2(P))\,
  \theta_2\w\xi^1
       \]
       \[\fl\hspace{15pt}
+(2\,P-3-2\,J_1\,J_2+2\,\D_2(J_2)+Q)\,\xi^1\w\sigma_{22},
\]
\noindent
where the functions
$J_1=-P_{tx}\,H^{-1}$ and $J_2=(H_t\,P_t-H\,P_{tt})\,H^{-1}\,(P_t)^{-2}$
are invariants of the symmetry group of an equation from $\mathcal{S}_2$,
the operators
\[
\D_1 = {{\partial}\over{\partial \xi^1}} = (P_t)^{-1}\,D_t,
\qquad
\D_2 = {{\partial}\over{\partial \xi^2}} = P_t\,H^{-1}\,D_x
\]
are invariant differentiations associated with $\xi^1$ and $\xi^2$. These
operators are defined by the identity
$dF = \D_1(F)\,\xi^1+\D_2(F)\,\xi^2$,
where $F=F(t,x)$ is an arbitrary function. The commutator identity for the invariant
differentiations has the form
\begin{equation}
\left[ \D_1, \D_2 \right] =
- J_1 \,\D_1
- J_2 \,\D_2.
\label{commutator_Q_2}
\end{equation}
We have $\D_1(P) = 1$, and, applying (\ref{commutator_Q_2}) to
$P$, we obtain the {\it syzygy}
\begin{equation}
J_1 = - \D_1(\D_2(P))-J_2\,\D_2(P).
\label{syzygy_Q_2}
\end{equation}
\noindent
If
$\D_2(P)\,\D_1(Q) \not = \D_2(Q)$, i.e.,
if $P_t\,Q_x \not = P_x\,Q_t$, then, applying (\ref{commutator_Q_2}) to $Q$
and using (\ref{syzygy_Q_2}), we have
\[
J_2 = \left(
\left[\D_1, \D_2\right] (Q)
- \D_1(Q)\,\D_1(\D_2(P))
\right)\,
(\D_2(P)\,\D_1(Q) - \D_2(Q))^{-1}.
\]
\noindent
Therefore, in this case the functions $P$ and $Q$ are a basis of differential
invariants of the symmetry group. But $P$ and $Q$ are not neccessary a basis
in the case of their functional dependence,
cf. \cite[th~2.3]{Ibragimov97}, \cite[\S~10.4.2]{Ibragimov99}.
To prove this statement, we consider the equation
\begin{equation}
u_{tx} = u_t + {{2\,(p(t)-1)}\over{q(t)\,(t+x)}}\,u_x
+ {{2}\over{q(t)\,(t+x)^2}}\,
\left(1-(p(t)-1)\,(t+x)\right)\,u
\label{counter_example}
\end{equation}
\noindent
with arbitrary functions $p(t)$ and $q(t)$ such that
$p^{\prime}(t) \not = 0$ and $q^{\prime}(t) \not = 0$. For this equation we
have
\[
P=p(t),
\quad
Q=q(t),
\quad
J_2 = -{{2}\over{q^{\prime}(t)\,(t+x)}}
- {{p^{\prime \prime}(t)}\over{(p^{\prime}(t))^2}}
- {{q^{\prime}(t)}\over{p^{\prime}(t)\,q(t)}},
\]
\noindent
$\D_1(P)= 1$,
$\D_2(P)= 0$,
$\D_1(Q)= q^{\prime}(t)\,(p^{\prime}(t))^{-1}$,
$\D_2(Q)= 0$, and by induction the only non-trivial higher order differential
invariants $\D^i_1(Q)$ depend on $t$. Since $J_{2,x} \not = 0 $, the function
$J_2$ is independent of $P$, $Q$, and all their invariant derivatives. Thus for
the whole subclass $\mathcal{S}_2$ we should take the functions $P$, $Q$, and
$J_2$ as a basis for the set of differential invariants of the symmetry group.
To construct all the other invariants, we apply $\D_1$ and
$\D_2$ to $P$, $Q$, and $J_2$. The commutator identity
(\ref{commutator_Q_2}) allows us to permute $\D_1$ and $\D_2$,
so we need only to deal with the invariants
$P_{jk}=\D_1^j(\D_2^k(P))$,
$Q_{jk}=\D_1^j(\D_2^k(Q))$, and
$J_{2,jk}=\D_1^j(\D_2^k(J_2))$, where $j \ge 0$, $k \ge 0$.

For $s \ge 0$ the $s$-th order  {\it classifying manifold} associated with
the lifted coframe $\bftheta$ and an open subset $U\subset\R^2$ is
\begin{equation}
\fl
\Cc^{(s)}(\bftheta,U) = \{(P_{jk}(t, x), Q_{jk}(t, x), J_{2,jk}(t, x))
\,\,\,\vert\,\,\,
0 \le j+k\le s, \,\,(t,x)\in U\}.
\label{manifold_Q_2}
\end{equation}
\noindent
Since all the functions $P_{jk}$, $Q_{jk}$, and $J_{2,jk}$  depend on two
variables $t$ and $x$, it follows that
$\rho_s = \dim \Cc^{(s)}(\bftheta,U) \le 2$ for all $s\ge 0$.
Let $r=\min \{s \,\,\vert\,\, \rho_s = \rho_{s+1} = \rho_{s+2} = ...\}$
be the {\it order of the coframe} $\bftheta$. Since $P_t \not = 0$, we have
$1\le \rho_0\le\rho_1\le\rho_2 \le ... \le 2$. In any case, $r+1 \le 2$.
Hence from Theorem 15.12 of \cite{Olver95} we see that
two linear hyperbolic equations (\ref{hyper}) from the subclass $\mathcal{S}_2$
are locally equivalent under a contact transformation if and only if
their second order classifying manifolds (\ref{manifold_Q_2}) locally overlap.

\vskip 5 pt
\noindent
{\bf Remark 1.}  A Lie pseudo-group is called structurally intransitive,
\cite{LisleReid}, if it is not isomorphic to any transitive Lie pseudo-group.
In \cite{Cartan4}, Cartan proved that a Lie pseudo-group is structurally
intransitive whenever it has essential invariants. An invariant of a Lie
pseudo-group with the structure equations
\[
d\omega^i=A^i_{\beta k}\,\pi^{\beta}\w\omega^k+T^i_{jk}\,\omega^j\w\omega^k
\]
\noindent is called {\it essential} if it is a first integral of the
{\it systatic system} $A^i_{\beta k}\,\omega^k$. From the structure equations
(\ref{SE_Q_2}) it follows that the systatic system for the sym\-met\-ry
pseudo-group of an equation from $\mathcal{S}_2$ is generated by the forms
$\xi^1$ and $\xi^2$. First integrals of these forms are arbitrary functions of
$t$ and $x$. Therefore, the invariants $P$, $Q$, $J_2$, and all the
non-constant derived invariants are essential. Thus the symmetry pseudo-group
of equation (\ref{hyper}) from the subclass $\mathcal{S}_2$ is structurally
intransitive, and the moving coframe method is applicable to finding
Maurer -- Cartan forms for differential equations with structurally
intransitive symmetry pseudo-groups, cf. \cite{LisleReid}.

\vskip 5 pt
\noindent
{\bf Remark 2.}  In \cite[th~1]{JMWS}, the following basis of invariants
for the symmetry group of equation (\ref{hyper}) is found:
$\{P, Q, J^1_3, J^2_3, J^3_3\}$, where
\[
J^1_3 = H^{-3}\,\left(K\,H_{tx}+H\,K_{tx}-H_tK_x-H_xK_t\right),
\]
\[
J^2_3 = H^{-9}\,\left(H\,K_x-K\,H_x\right)^2\,
\left(H\,K\,H_{tt}-H^2K_{tt}-3\,K\,H^2_t+3\,H\,H_tK_t\right),
\]
\[
J^3_3 = H^{-9}\,\left(H\,K_t-K\,H_t\right)^2\,
\left(H\,K\,H_{xx}-H^2K_{xx}-3\,K\,H^2_x+3\,H\,H_xK_x\right).
\]
\noindent Using (\ref{syzygy_Q_2}), we have the following
expressions for invariants $J^1_3$, $J^2_3$, and $J^3_3$ in terms of $P$,
$Q$, $J_2$, and their invariant derivatives:
\[
J^1_3 = 2\,P\,Q+\D_1(\D_2(P)) + J_2\,\D_2(P),
\]
\[
J^2_3 = J_2\, \left(\D_2(P)\right)^2,
\]
\[
J^3_3 = \D_2(P)\,\left(\D_1(\D_2(P))
+J_2\,\D_2(P)\right) -\D_2(\D_2(P)).
\]
The following operators of invariant differentiation are found in \cite{JMWS}:
\[
\wt{X}_1 = H^{-3}\,\left(H\,K_x-K\,H_x\right)\,D_t,\qquad
\wt{X}_2 = H^2\,\left(H\,K_x-K\,H_x\right)^{-1}\,D_x.
\]
\noindent
We have $\wt{X}_1 = \D_2(P)\,\D_1$ and
$\wt{X}_2 = \D_2(P)^{-1}\,\D_2$. Then in the case $\D_2(P) \equiv 0 \equiv P_x$ 
the operator $\wt{X}_2$ is not
defined, while $\wt{X}_1$ is trivial, $J^1_3 = 2\,P\,Q$, $J^2_3 = 0$, and
$J^3_3 = 0$. Therefore, the functions $P$, $Q$, $J^1_3$, $J^2_3$, and $J^3_3$
are not a basis of invariants of symmetry group for equation
(\ref{counter_example}).

\vskip 5 pt
\noindent
{\bf Remark 3.}  In the theorem of \cite{Ibragimov2004}, two sets of functions
are stated to be bases for invariants of symmetry groups of equations (\ref{hyper}):
the first set consists of functions $P$, $Q$, $I = P_tP_xH^{-1}$,
$\wt{Q} = (\ln \vert\,K\,\vert\,)_{tx}K^{-1}$, and the second set consists
of the functions $P$, $Q$, $I$, and $-J_2$. The operators of invariant
differentiation are taken in the form $\Dc_1 = P_t^{-1}D_t$ and
$\Dc_2 = P_x^{-1}D_x$. We have $I = \D_2(P)$, therefore the function $I$ can
be excluded from both sets. Also we have
$\wt{Q} = Q\,P^{-1}+ J_2 \, \D_2(P)\,P^{-2} +\D_1(\D_2(P)) \, P^{-2}
- \D_2(P)\,P^{-3}$, $\Dc_1 = \D_1$, and $\Dc_2 = (\D_2(P))^{-1}\,\D_2$.
Therefore, in the case $P_x = 0 = \D_2(P)$ we have $I=0$ and
$\wt{Q} = Q\,P^{-1}$, so the functions $P$, $Q$, $I$, and
$\wt{Q}$ are not a basis of invariants for the symmetry group of equation
(\ref{counter_example}).

The function  $J_2$  and the operator $\Dc_1$  are not defined
when $P_t \equiv 0$ (for example of this case we take the Moutard equation
$u_{tx} = U(t,x)\,u$). So the second set of functions is not a basis of
invariants of symmetry groups for the {\it whole} class (\ref{hyper}).

\vskip 5 pt

Now we return to the case $P_t \equiv 0$. Then the torsion coefficients in the
structure equations (\ref{SE_P}) are independent of the group parameters,
while $d P = P_x\,b^1_1\,H^{-1}\,\xi^2$. We denote by $\mathcal{S}_3$ the
subclass of equations (\ref{hyper}) such that $P_t \equiv 0$, $P_x \not = 0$.
This subclass is not empty, since, for example, the equation
$u_{tx} = x^2\,u_x+ u$ belongs to $\mathcal{S}_3$. For an equation from
$\mathcal{S}_3$ we normalize $b^1_1 = H\,P_x^{-1}$.
After absorption of torsion and prolongation, we obtain the involutive lifted
coframe
$\bftheta = \{\theta_0, \theta_1, \theta_2, \xi^1,
\xi^2, \sigma_{11}, \sigma_{22}, \eta_1, \eta_2, \eta_3\}$
with the structure equations
\[\fl
d\theta_0=
 \eta_1\w\theta_0
+\xi^1\w\theta_1
+\xi^2\w\theta_2,
\]
\[\fl
d\theta_1=
 \eta_1\w\theta_1
-P\,\theta_0\w\xi^2
-L\,\theta_1\w\xi^2
+\xi^1\w\sigma_{11},
\]
\[\fl
d\theta_2=
\eta_1\w\theta_2
-\theta_0\w\xi^1
+L\,\theta_2\w\xi^2
+ \xi^2\w\sigma_{22},
\]
\[\fl
d\xi^1=
L\,\xi^1\w\xi^2,
\]
\[\fl
d\xi^2=0,
\]
\[\fl
d\sigma_{11}=
 \eta_1\w\sigma_{11}
+\eta_2\w\xi^1
+ (Q+1-2\,P)\,\theta_1\w\xi^2
+2\,L\,\xi^2\w\sigma_{11},
\]
\[\fl
d\sigma_{22}=
 \eta_1\w\sigma_{22}
+\eta_3\w\xi^2
+(P-2+Q)\,\theta_2\w\xi^1,
\]
\[\fl
d\eta_1=
(P-1)\,\xi^1\w\xi^2,
\]
\[\fl
d\eta_2=
\pi_1\w\xi^1
-\eta_1\w\eta_2
-3\,L\,\eta_2\w\xi^2
-\D_1(Q)\,\theta_1\w\xi^2
+(3\,Q-3\,P+2)\,\xi^2\w\sigma_{11},
\]
\[\fl
d\eta_3=
\pi_2\w\xi^2
+\eta_1\w\eta_3
-(4\,L+1-2\,P\,L-2\,Q\,L+\D_2(Q))\,\theta_2\w\xi^1
      \]
      \[\fl\hskip 15 pt
+(Q-3+2\,P)\,\xi^1\w\sigma_{22},
\]
\noindent where the function
$L=\left(H\,P_{xx} - H_x\,P_x\right)\,(P_x)^{-2}\,H^{-1}$ is an invariant of
the symmetry group, and the operators of invariant differentiation are
$\D_1 = P_x\,H^{-1}\,D_t$ and $\D_2 = (P_x)^{-1}\,D_x$.
We have $\D_1(P)= 0$, $\D_2(P)=1$, and
\begin{equation}
\left[\D_1, \D_2\right] = L\, \D_1.
\label{commutator_Q_3}
\end{equation}
In the case $\D_1(Q) \not = 0$ we apply (\ref{commutator_Q_3}) to $Q$ and
obtain $L=\left[\D_1,\D_2\right](Q)\,(\D_1(Q))^{-1}$. Therefore, in this case
the functions $P$ and $Q$ are a basis for the set of differential invariants
of the symmetry group. But if $\D_1(Q)=0$, then the functions $P$ and $Q$ are
not neccessary a basis. For example, consider the equation
\[
u_{tx} = - {{2\,(p(x)-1)}\over{q(x)\,(t+x)}}\,u_t + u_x
+ {{2}\over{q(x)\,(t+x)^2}}\,
\left(p(x)+(p(x)-1)\,(t+x)\right)\,u
\]
\noindent
where $p(x)$ and $q(x)$ are arbitrary functions such that
$p^{\prime}(x) \not = 0$ and $q^{\prime}(x) \not = 0$.
This equation has the following invariants:
\[
P=p(x), \quad Q=q(x), \quad
L = {{2}\over{p^{\prime}(x)\,(t+x)}}
+ {{p^{\prime \prime}(x)}\over{(p^{\prime}(x))^2}}
+ {{q^{\prime}(x)}\over{p^{\prime}(x)\,q(x)}}.
\]
\noindent
We have $\D_1(Q)= 0$, $\D_2(Q)= q^{\prime}(x)\,(p^{\prime}(x))^{-1}$, and by
induction the only non-trivial higher order differential invariants $\D^i_2(Q)$
depend on $x$. Since $L_{t} \not = 0 $, the function $L$ is independent of $P$,
$Q$, and all their invariant derivatives. Thus for the whole subclass
$\mathcal{S}_3$ we should take the functions $P$, $Q$, and $L$ as a basis for
the set of differential invariants of symmetry group. The $s$-th order
classifying manifold associated with the coframe $\bftheta$ and an open subset
$U \in \R^2$ can be taken in the form
\begin{equation}
\fl
\Cc^{(s)}(\bftheta, U) = \{(P(x), Q_{jk}(t, x), L_{jk}(t, x))
\,\,\,\vert\,\,\,
0 \le j+k \le s, \,\,(t,x)\in U\}
\label{manifold_Q_3}
\end{equation}
\noindent
with $Q_{jk}=\D_1^j(\D_2^k(Q))$ and $L_{jk}=\D_1^j(\D_2^k(L))$.
Then two equations from $\mathcal{S}_3$ are equivalent under a contact
transformation if and only if their second order classifying manifolds
(\ref{manifold_Q_3}) are (locally) overlap.

\vskip 5 pt

Now we consider the case $P \equiv const$. Then we have
$dQ = Q_t\,(b^1_1)^{-1}\,\xi^1+Q_x\,b^1_1\,H^{-1}\,\xi^2$. We denote by
$\mathcal{S}_4$ the subclass of equations (\ref{hyper}) such that 
$P \equiv const.$, $Q_t \not = 0$. This subclass is not empty, since, for
example, the equation $u_{tx} = (t-x)^3\,u_x + (t-x)^2\,u$ belongs to
$\mathcal{S}_4$. For an equation from $\mathcal{S}_4$ we normalize
$b^1_1 = Q_t$. Then after absorption of torsion and prolongation we have the
involutive lifted coframe $\bftheta = \{\theta_0, \theta_1, \theta_2, \xi^1,
\xi^2, \sigma_{11}, \sigma_{22}, \eta_1, \eta_2, \eta_3\}$ with the structure
equations
\[\fl
d\theta_0=
\eta_1\w\theta_0
+\xi^1\w\theta_1
+\xi^2\w\theta_2,
\]
\[\fl
d\theta_1=
\eta_1\w\theta_1
-P\,\theta_0\w\xi^2
-M_2\,\theta_1\w\xi^1
-M_1\,\theta_1\w\xi^2
+\xi^1\w\sigma_{11},
\]
\[\fl
d\theta_2=
\eta_1\w\theta_2
-\theta_0\w\xi^1
+M_2\,\theta_2\w\xi^1
+M_1\,\theta_2\w\xi^2
+\xi^2\w\sigma_{22},
\]
\[\fl
d\xi^1=M_1\,\xi^1\w\xi^2,
\]
\[\fl
d\xi^2=
M_2\,\xi^1\w\xi^2,
\]
\[\fl
d\sigma_{11}=
\eta_1\w\sigma_{11}
+\eta_2\w\xi^1
+(Q+1-2\,P)\,\theta_1\w\xi^2
+2\,M_1\,\xi^2\w\sigma_{11},
\]
\[\fl
d\sigma_{22}=
\eta_1\w\sigma_{22}
+\eta_3\w\xi^2
+(P-2+Q)\,\theta_2\w\xi^1
-2\,M_2\,\xi^1\w\sigma_{22},
\]
\[\fl
d\eta_1=
(P-1)\,\xi^1\w\xi^2,
\]
\[\fl
d\eta_2=
\pi_1\w\xi^1
+\eta_1\w\eta_2
-3\,M_1\,\eta_2\w\xi^2
-(1+2\,M_2+2\,Q\,M_2-4\,P\,M_2)\,\theta_1\w\xi^2
      \]
      \[\fl\hskip 15 pt
+(Q-2\,M_1M_2-3\,P-2\,\D_1(M_1)+2)\,
\xi^2\w\sigma_{11},
\]
\[\fl
d\eta_3=
\pi_2\w\xi^2
+\eta_1\w\eta_3
+3\,M_2\,\eta_3\w\xi^1
-(4\,M_1-2\,M_1P-2\,M_1Q+\D_2(Q))\,
\theta_2\w\xi^1
      \]
      \[\fl\hskip 15 pt
+(2\,M_1M_2+2\,P-3+2\,\D_1(M_1)+3\,Q)\,\xi^1\w\sigma_{22},
\]
\noindent
where the functions $M_1=-Q_{tx}\,H^{-1}$ and
$M_2=\left(H_tQ_t-H\,Q_{tt}\right)\,H^{-1}\,Q_t^{-2}$ are invariants of the
symmetry group, and the operators of invariant differentiation are
$\D_1 = Q_t^{-1}\,D_t$ and $\D_2 = Q_t\,H^{-1}\,D_x$. We have
$\left[\D_1,\D_2\right] = -M_1\,\D_1-M_2\,\D_2$. Since $\D_1(Q)=1$, then,
applying the commutator identity to $Q$, we have the syzygy
$M_1 = -\D_1(\D_2(Q)) - M_2\,\D_2(Q)$. The functions $Q$ and $M_2$ are a
basis for the set of all invariants of the symmetry group of an equation from
$\mathcal{S}_4$. We take the $s$-th order classifying manifold associated with
the coframe $\bftheta$ and an open subset $U \in \R^2$ in the form
\begin{equation}
\fl
\Cc^{(s)}(\bftheta, U) = \{(P, Q_{jk}(t, x), M_{2,jk}(t, x))
\,\,\,\vert\,\,\,
0 \le j+k \le s, \,\,(t,x)\in U\}
\label{manifold_Q_4}
\end{equation}
\noindent
with $Q_{jk}=\D_1^j(\D_2^k(Q))$ and $M_{2,jk}=\D_1^j(\D_2^k(M_2))$. Then two
equations from $\mathcal{S}_4$ are equivalent under a contact transformation
if and only if their second order classifying manifolds (\ref{manifold_Q_4})
are (locally) overlap.

\vskip 5 pt

Next we denote by $\mathcal{S}_5$ the subclass of equations (\ref{hyper})
such that $P \equiv const.$, $Q_t \equiv 0$, $Q_x \not = 0$. This subclass is
not empty,  since, for example, the equation
\[
u_{tx} =
-{\frac {2\,(\lambda-1)} {q(x)\,(t+x)}}\,u_t + u_x
+ {\frac {2\,(\lambda+(\lambda-1)\,(t+x))} {q(x)\,(t+x)^2}}\,u
\]
\noindent has the invariants $P = \lambda \equiv const.$, $Q = q(x)$, and
belongs to $\mathcal{S}_5$. For an equation from $\mathcal{S}_5$ we normalize
$b^1_1 = H\,Q_x^{-1}$. Then after absorption of torsion and prolongation we
have the involutive lifted coframe
$\bftheta = \{\theta_0, \theta_1, \theta_2, \xi^1,
\xi^2, \sigma_{11}, \sigma_{22}, \eta_1, \eta_2, \eta_3\}$
with the structure equations
\[\fl
d\theta_0=
\eta_1\w\theta_0
+\xi^1\w\theta_1
+\xi^2\w\theta_2,
\]
\[\fl
d\theta_1=
\eta_1\w\theta_1
-P\,\theta_0\w\xi^2
-N\,\theta_1\w\xi^2
+\xi^1\w\sigma_{11},
\]
\[\fl
d\theta_2=
\eta_1\w\theta_2
-\theta_0\w\xi^1
+N\,\theta_2\w\xi^2
+\xi^2\w\sigma_{22},
\]
\[\fl
d\xi^1=
N\,\xi^1\w\xi^2,
\]
\[\fl
d\xi^2=0,
\]
\[\fl
d\sigma_{11}=
\eta_1\w\sigma_{11}
+\eta_2\w\xi^1
+(Q+1-2\,P)\,\theta_1\w\xi^2
+2\,N\,\xi^2\w\sigma_{11},
\]
\[\fl
d\sigma_{22}=
\eta_1\w\sigma_{22}
+\eta_3\w\xi^2
+(P-2+Q)\,\theta_2\w\xi^1,
\]
\[\fl
d\eta_1=
(P-1)\,\xi^1\w\xi^2,
\]
\[\fl
d\eta_2=
\pi_1\w\xi^1
+\eta_1\w\eta_2
-3\,N\,\eta_2\w\xi^2
+(2-3\,P+3\,Q)\,\xi^2\w\sigma_{11},
\]
\[\fl
d\eta_3=
\pi_2\w\xi^2
+\eta_1\w\eta_3
+(2\,N\,(P+Q-2)-1)\,\theta_2\w\xi^1
+(2\,P+Q-3)\,\xi^1\w\sigma_{22},
\]
\noindent
where the function $N=(H\,Q_{xx}-H_x\,Q_x)\,H^{-1}\,Q_x^{-2}$ is an invariant
of the symmetry group, and the operators of invariant differentiation are
$\D_1=Q_x\,H^{-1}\,D_t$ and $\D_2=Q_x^{-1}\,D_x$. We have
$\left[\D_1, \D_2\right] = - N\, \D_1$, $\D_1(Q) = 0$, and $\D_2(Q) = 1$.
The functions $Q$ and $N$ are a basis for the set of all invariants of the
symmetry group of an equation from $\mathcal{S}_5$. We take the $s$-th order
classifying manifold associated with the coframe $\bftheta$ and an open subset
$U \in \R^2$ in the form
\begin{equation}
\fl
\Cc^{(s)}(\bftheta, U) = \{ (P, Q(x),
\D_1^j(\D_2^k(N))(t, x))
\,\,\,\vert\,\,\,
0 \le j+k \le s, \,\,(t,x)\in U\}.
\label{manifold_Q_5}
\end{equation}
\noindent
Then two equations from $\mathcal{S}_5$ are equivalent under a contact
transformation if and only if their second order classifying manifolds
(\ref{manifold_Q_5}) are (locally) overlap.

\vskip 5 pt

Finally, we denote by $\mathcal{S}_6$ the subclass of equations (\ref{hyper})
such that $P \equiv const.$, $Q \equiv const.$ This subclass is not empty,
since, for example, the equation
\begin{equation}
u_{tx} = -t\,u_t -\lambda\,x\,u_x -\lambda\,t\,x\, u
\label{S_6_1}
\end{equation}
\noindent
has the invariants $P=\lambda$ and $Q=0$, while the Euler - Poisson equation
\begin{equation}
u_{tx} = 2\,\mu^{-1}\,(t+x)^{-1}\,u_t
+2\,\lambda\,\mu^{-1}\,(t+x)^{-1}\,u_x
- 4\,\lambda\,\mu^{-2}\,(t+x)^{-2}\,u
\label{S_6_2}
\end{equation}
\noindent
has the invariants $P = \lambda$ and $Q = \mu$, \cite[\S~9.2]{Ovsiannikov}.
For an equation from $\mathcal{S}_6$ after absorption of torsion and
prolongation we have the involutive lifted coframe
$\bftheta = \{\theta_0, \theta_1, \theta_2, \xi^1,
\xi^2, \sigma_{11}, \sigma_{22}, \eta_1, \eta_2, \eta_3, \eta_4\}$
with the structure equations
\[
d\theta_0=
\eta_1\w\theta_0
+\xi^1\w\theta_1
+\xi^2\w\theta_2,
\]
\[
d\theta_1=
\eta_2\w\theta_1
-P\,\theta_0\w\xi^2
+\xi^1\w\sigma_{11},
\]
\[
d\theta_2=
(2\,\eta_1
-\eta_2)\w\theta_2
-\theta_0\w\xi^1
+\xi^2\w\sigma_{22},
\]
\[
d\xi^1=
(\eta_1-\eta_2)\w\xi^1,
\]
\[
d\xi^2=
(\eta_2 -\eta_1)\w\xi^2,
\]
\[
d\sigma_{11}=
(2\,\eta_2-\eta_1)\w\sigma_{11}
+\eta_3\w\xi^1
+(Q+1-2\,P)\,\theta_1\w\xi^2,
\]
\[
d\sigma_{22}=
(3\,\eta_1-2\,\eta_2)\w\sigma_{22}
+\eta_4\w\xi^2
+(P-2+Q)\,\theta_2\w\xi^1,
\]
\[
d\eta_1=
(P-1)\,\xi^1\w\xi^2,
\]
\[
d\eta_2=
(P-Q-1)\,\xi^1\w\xi^2,
\]
\[
d\eta_3=
\pi_1\w\xi^1
- (2\,\eta_1-3\,\eta_2)\w\eta_3
+(3\,(Q-P)+2)\,\xi^2\w\sigma_{11},
\]
\[
d\eta_4=
\pi_2\w\xi^2
+(4\,\eta_1-3\,\eta_2)\w\eta_4
+(3\,(Q-1)+2\,P)\,\xi^1\w\sigma_{22}.
\]
\noindent All the invariants of the symmetry group for an equation from
$\mathcal{S}_6$ are constants, and the classifying manifold is a point.
Thus an equation from $\mathcal{S}_6$ is equivalent to one of equations
(\ref{S_6_1}) or (\ref{S_6_2}) with the same values of $P$ and $Q$,
\cite[\S~9.2]{Ovsiannikov}.

\vskip 5 pt
The results of the above calculations are summarized in the following statement:

\vskip 5 pt
\noindent{\bf Theorem.}\,\,
{\it The class of linear hyperbolic equations (\ref{hyper}) is divided into
the six subclasses} $\mathcal{S}_1$, $\mathcal{S}_2$, ..., $\mathcal{S}_6$
{\it invariant under an action of the pseudo-group of contact transformations:}

$\mathcal{S}_1$ {\it consists of all equations (\ref{hyper}) such that}
$H \equiv 0$ {\it and} $K \equiv 0$;

$\mathcal{S}_2$ {\it consists of all equations (\ref{hyper}) such that}
$P_t \not= 0$;

$\mathcal{S}_3$ {\it consists of all equations (\ref{hyper}) such that}
$P_t \equiv 0$ {\it and}
$P_x \not = 0$;

$\mathcal{S}_4$ {\it consists of all equations (\ref{hyper}) such that}
$P \equiv const.$
{\it and}
$Q_t \not = 0$;

$\mathcal{S}_5$ {\it consists of all equations (\ref{hyper}) such that}
$P \equiv const.$,
$Q_t \equiv 0$, {\it and}
$Q_x \not = 0$;

$\mathcal{S}_6$ {\it consists of all equations (\ref{hyper}) such that}
$P \equiv const.$
{\it and}
$Q \equiv const.$

{\it Every equation from the subclass} $\mathcal{S}_1$ {\it is locally equivalent to
the linear wave equation} $u_{tx} = 0$.

{\it Every equation from the subclass} $\mathcal{S}_6$ {\it is locally
equivalent to either equation (\ref{S_6_1}) when} $Q = 0$ {\it or to the
equation (\ref{S_6_2}) when} $Q \not = 0$.

{\it For the subclass}  $\mathcal{S}_2$, {\it the basic invariants are}  $P$,
$Q$, {\it and}
$J_2$,
{\it the operators of  invariant differentiation are}
$\D_1 = P_t^{-1} D_t$ {\it and}
$\D_2 = P_t H^{-1} D_x$.

{\it For the subclass}  $\mathcal{S}_3$, {\it the basic invariants are}
$P$, $Q$, {\it and} $L$, {\it the operators of  invariant differentiation are}
$\D_1 = P_x H^{-1} D_t$ {\it and} $\D_2 = P_x^{-1} D_x$.

{\it For the subclass}  $\mathcal{S}_4$, {\it the basic invariants are}
$Q$, $M_1$, {\it and}
$M_2$,
{\it the operators of  invariant differentiation are}
$\D_1 = Q_t^{-1} D_t$ {\it and}
$\D_2 = Q_t H^{-1} D_x$.

{\it For the subclass} $\mathcal{S}_5$, {\it the basic invariants are}
$Q$ {\it and} $N$,
{\it the operators of  invariant differentiation are}
$\D_1 = Q_x H^{-1} D_t$ {\it and}
$\D_2 = Q_x^{-1} D_x$.

{\it Two equations from one of the subclasses} $\mathcal{S}_2$, $\mathcal{S}_3$,
$\mathcal{S}_4$, {\it or}
$\mathcal{S}_5$ {\it are locally equivalent to each other if and only if the
classifying manifolds
(\ref{manifold_Q_2}),
(\ref{manifold_Q_3}),
(\ref{manifold_Q_4}),
or
(\ref{manifold_Q_5}) for these equations locally overlap.}

\section*{Conclusion}
In this paper, the moving coframe method of \cite{FO} is applied to
the local equivalence problem for the class of linear second-order hyperbolic
equations in two independent variables under an action of the pseudo-group of
contact transformations. The class is divided into the six invariant
subclasses. For all the subclasses, the Maurer - Cartan forms for symmetry
groups, the bases of differential invariants and the invariant differentiation
operators are found. This allowed to solve the equivalence problem for
the whole class of linear hyperbolic equations. It is shown that the moving
coframe method is applicable to structurally intransitive symmetry groups. The
method uses linear algebra and differentiation operations only and does not
require analysing over-determined systems of partial differential equation or
using procedures of integration.

\end{document}